\documentclass[a4paper,12pt]{article}
\usepackage[utf8]{inputenc}
\usepackage[T1]{fontenc}
\usepackage[english]{babel}
\usepackage{amsfonts,amsbsy,bm,euscript,mathrsfs}
\usepackage[tbtags]{amsmath}
\usepackage{authblk}
\usepackage{amssymb,stmaryrd,faktor}
\usepackage{graphicx}
\usepackage{caption}
\usepackage[title,titletoc]{appendix}
\usepackage{titlesec}
\usepackage[bookmarks=true,colorlinks=true,linkcolor=black,citecolor=blue,urlcolor=blue,bookmarksnumbered]{hyperref}
\usepackage{overpic}
\usepackage{upgreek}
\usepackage{mathabx} 
\usepackage{soul}
\usepackage[dvipsnames]{xcolor}

\usepackage{setspace}

\usepackage{dsfont}
\usepackage{collref}
\usepackage{lmodern}
\usepackage{mathrsfs}
\usepackage{mathtools}
\usepackage{fixmath}
\usepackage{cancel}
\usepackage{cases}

\usepackage{bbm}
\usepackage{braket}
\usepackage{slashed}

\usepackage{booktabs}
\usepackage{subfig}
\usepackage{tikz}
\usetikzlibrary{plotmarks,calc,decorations,decorations.markings,decorations.pathmorphing}
\usetikzlibrary{shapes,arrows.meta,automata,positioning}
\usepackage{tikz-3dplot}
\usepackage{circuitikz}
\usepackage{slashed}

\usepackage[usenames,dvipsnames]{pstricks}
\usepackage{pstricks-add}
\usepackage{epsfig}
\usepackage{pst-grad} 
\usepackage{pst-plot} 
\usepackage[space]{grffile} 
\usepackage{etoolbox} 
\makeatletter 
\patchcmd\Gread@eps{\@inputcheck#1 }{\@inputcheck"#1"\relax}{}{}
\makeatother

\usepackage{blkarray}
\usepackage{comment}

\numberwithin{equation}{section}

\usepackage{titlesec}

\newcommand{\tx}[1]{\text {#1}}
\newcommand{\cl}[1]{\mathcal {#1}}
\newcommand{\sr}[1]{\mathscr {#1}}
\newcommand{\bb}[1]{\mathbb {#1}}
\newcommand{\fk}[1]{\mathfrak {#1}}

\newcommand{\ol}[1]{\overline {#1}}

\def\cpn{{$ \mathbb{CP}^{n-1} $}}

\newcommand{\nd}{\noindent}

\makeatletter
\newsavebox\BoxA
\newsavebox\BoxB
\newlength\LengthA

\newcommand*\obr[2][0.75]{%
	\sbox{\BoxA}{$\m@th#2$}%
	\setbox\BoxB\null
	\ht\BoxB=\ht\BoxA%
	\dp\BoxB=\dp\BoxA%
	\wd\BoxB=#1\wd\BoxA
	\sbox\BoxB{$\m@th\overline{\copy\BoxB}$}
	\setlength\LengthA{\the\wd\BoxA}
	\addtolength\LengthA{-\the\wd\BoxB}%
	\ifdim\wd\BoxB<\wd\BoxA%
	\rlap{\hskip 0.5\LengthA\usebox\BoxB}{\usebox\BoxA}%
	\else
	\hskip -0.5\LengthA\rlap{\usebox\BoxA}{\hskip 0.5\LengthA\usebox\BoxB}%
	\fi}
\makeatother

\begin{document}
	\title{Superdeformed $\mathbb{CP}$ $\sigma$-model equivalence} 
	\author[1,2]{Anton Pribytok\footnote{antonspribitoks@bimsa.cn, a.pribytok@gmail.com}} 
	\affil[1]{\small Yanqi Lake Beijing Institute of Mathematical Sciences and Applications (BIMSA), Huairou District, Beijing 101408, P. R. China}
	\affil[2]{\small Steklov Mathematical Institute of Russian Academy of Sciences, \protect\\ Gubkina str. 8, 119991 Moscow, Russia} 
	\date{}

\maketitle

\abstract{\nd We find the novel class of the supersymmetric deformation of the $\bb{CP}^{1}$ $\sigma$-model and its equivalence with the generalised chiral Gross-Neveu. This construction allows the use of field-theoretic techniques and particularly the study of renormalisability and $\beta$-function. Provided approach is useful in finding conformal limits and establishes relation between chiral (GN) and sigma model description (geometric), which is explicitly demonstrated for the case of $ \mathbb{R} \times S^{1} $/Super-Thirring models. We also provide discussion on its emergence in $\mathcal{N}=2$ Liouville and 4-dim Chern-Simons theory.}
	
	\vspace{4.00mm}
	\nd Prepared as a contribution to "\textit{Supersymmetries and Quantum Symmetries – SQS'24}". Based on recent progress \cite{Bykov:2023klm}, \cite{Four-dim_CS_OD}.


	\section{ $\bb{CP}^{n-1} $ sigma model from chiral formalism} 
	In this work we address the construction of new deformations of sigma models from the chiral field-theoretic formalism. More specifically, we exploit Gross-Neveu model framework to build a chiral analogue of the supersymmetric deformation of the $\bb{CP}^{1}$ sigma model, which by now itself lacked a Lagrangian description. Our approach allows one to implement all field-theoretic techniques for computation of observables, study new model properties and other, which is implicit or inaccessible from the worldsheet perspective.
	
	To demonstrate the core structure of the approach, we recall the relation of the Gross-Neveu (GN) and $\bb{CP}^{n-1}$ sigma model. The chiral $N$-flavor GN model \cite{Anselm,GrossNeveu} can be defined as 
	\begin{equation}\label{GN}
		\sr{L}_{\tx{GN}} = \sum_{i=1}^{N}\ol{\Psi}_{i} \slashed{D} \Psi_{i}  + \varkappa \sum_{i,j=1}^{N} \left[ \ol{\Psi}_{i} \Gamma_{+} \Psi_{i}\right] \left[ \ol{\Psi}_{j} \Gamma_{-} \Psi_{j}\right] \,,
	\end{equation}
	where $\Psi$ is a Dirac spinor, $\varkappa$ is a coupling constant and projectors $\Gamma_{\pm} = (1\pm\gamma_{5})/2$. By taking into account Dirac spinor decomposition into Weyl components
	\begin{equation}
		\Psi_{a} = 
		\begin{pmatrix}
			U_{a} \\
			\bar{V}_{a}
		\end{pmatrix}
		\qquad a = \overline{1,n}
	\end{equation}
	where for now we assume commutativity at the level of $U,V$ components. In order to establish an equivalence with the sigma model formulation, one can recast \eqref{GN} in $ U, V $ 
	\begin{equation}
		\begin{array}{c}
			\mathscr{L} = \sum_{a=1}^n\left(V_a \cdot \bar{D} U_a+\bar{U}_a \cdot D \bar{V}_a\right)+\kappa\left(\sum_{a=1}^n\left|U_a\right|^2\right)\left(\sum_{b=1}^n\left|V_b\right|^2\right) \\[2ex]
			\bar{D} U_{a} = \partial_{\bar{z}} U_{a} - i \bar{A} U_{a}
		\end{array} \,,
	\end{equation}
	which invariant under $U(1)$ type complex transform $U_{a} \rightarrow \lambda U_{a} \,, V_{a} \rightarrow \lambda^{-1} V_{a} \,, \lambda \in \bb{C}^{*}$. Since $ \sr{L}(U,V) $ is quadratic in field variables and $ U \leftrightarrow \bar{V}, \, z \leftrightarrow \bar{z} $, one can eliminate either through equations of motion and obtain 
	\begin{equation}\label{CPn_nf}
		\sr{L} = \dfrac{1}{\kappa} \dfrac{| \bar{D} U |^{2}}{| U |^{2}}
	\end{equation}
	that is a form of the $\bb{CP}^{n-1}$ model. At the same time recalling local projective invariance 
	\begin{equation}\label{Hopf_Fibration_Condition}
		U_{a} \rightarrow \lambda U_{a} \quad \xrightarrow{\,\,\,\, Hopf \, gauge \,\,\,\,} \quad \sum_{a=1}^{n} \bar{U}_{a} U_{a} = 1
	\end{equation}
	corresponding to the Hopf fibration $S^{2n-1} \rightarrow \bb{CP}^{n-1}$, which is apparent due to the residual $U(1)$. By exploiting \eqref{CPn_nf} and \eqref{Hopf_Fibration_Condition}, it can be shown that difference from the {\cpn} model results in 
	\begin{equation}
		\sr{L} - \sr{L}^{\,\tx{\cpn}} = \dfrac{1}{2 \kappa} \left( | \bar{D} U |^{2} - | D U |^{2} \right). 
	\end{equation}
	Hence the introduced Lagrangian differs by a topological term, \textit{i.e.} by the Fubini-Study form 
	\begin{equation}
		\sum_{a=1}^{n} \varepsilon_{\mu\nu} D_{\mu} \bar{U}_{a} D_{\nu} U_{a} \mapsto \Omega_{FB} = i \sum_{a=1}^{n} d\bar{U}_{a} \wedge d U_{a}
	\end{equation} 
	and classically\footnote{Addition of topological terms at the quantum level requires separate investigation (incl. $ \theta \int d^{2} z \, F_{z\bar{z}} $)} models agree \cite{BykovGN,BykovSUSY}. 
	
	\paragraph{Supersymmetrisation.} There are several possibilities to couple fermions to the bosonic $ \bb{CP}^{n-1} $ model. One suitable choice would be supersymmetric coupling, where one can start with $\beta\gamma$-system with $ \Phi = T^{*} C^{n} $ space (in chiral formalism -- phase space) and supersymmetrically couple fermions 
	\begin{equation}\label{Super_Beta_Gamma}
		\sr{L}_{\beta\gamma} = V \bar{\partial} U + \bar{U} \partial \bar{V} \,\, \Rightarrow \,\, \sr{L}_{\beta\gamma}^{\tx{\tiny Super}} = \sr{V}\bar{\partial}\sr{U} + \bar{\sr{U}}\partial\bar{\sr{V}} \qquad
		\sr{U} = 
		\begin{pmatrix}
			U \\
			C
		\end{pmatrix} \,,
		\sr{V} = 
		\begin{pmatrix}
			V & B
		\end{pmatrix} \,,
	\end{equation} 
	where $\sr{U, V}$ are super-doublets with $U, V$ to be bosonic and $B, C$ fermionic. The target space symmetry is given by $\sr{U} \rightarrow g \cdot \sr{U} $, $ \sr{V} \rightarrow \sr{V} \cdot g^{-1} $, $ g \in GL(n|n) $. At the worldsheet level $ \sr{L}_{\beta\gamma}^{\tx{\tiny Super}} $ also becomes invariant under supersymmetry transformations \cite{FMS} 
	\begin{equation}\label{Worldsheet_Transformations}
		\delta U=\epsilon_1 C, \quad \delta B=-\epsilon_1 V, \quad \delta C=-\epsilon_2 \partial U, \quad \delta V=\epsilon_2 \partial B \,,
	\end{equation} 
	which together with anti-holomorphic part closes to the $ \cl{N}=(2,2) $ superalgebra. As it can be noticed, transition to the {\cpn} model requires covariantisation 
	\begin{equation}\label{Free_SST}
		\sr{L} = (V \bar{\cl{D}} U + \bar{U} \cl{D} \bar{V}) + (B \bar{\cl{D}} C + \bar{C} \cl{D} \bar{B}) \qquad \bar{\cl{D}} = \bar{\partial} + i \bar{A} \,,
	\end{equation} 
	hence supersymmetric invariance is no longer manifest and subject to constraints \\$ \delta (B \cdot U) = -\epsilon_{1} \left( V\cdot U + B \cdot C \right) $ (involves gauging $ \bb{C}^{\times} \subset GL(1|1) $).
	
	Further step requires to make the theory interacting, which can be achieved by introducing a current-current type interaction. More specifically, the extended $ \bb{CP}^{n-1} $ Lagrangian can now be extended from \eqref{Free_SST} to
	\begin{equation}\label{Super-CPn}
		\cl{S} = 2 \int d^{2}z \, \sr{L}_{\fk{s}} \qquad \sr{L}_{\fk{s}} = \sr{V}\bar{\sr{D}}\sr{U} + \bar{\sr{U}}\sr{D}\bar{\sr{V}} + \frac{\varkappa}{2} \, \tx{Tr} \left[ \cl{J} \, \bar{\cl{J}} \right] \,.
	\end{equation}
	Here we have used the notation \eqref{Super_Beta_Gamma} for $\sr{U, V}$, $\varkappa$ is the coupling constant, currents are bilinears in fields $ \cl{J} = U \otimes V - C \otimes B $ (Kac-Moody currents) and bar implies conjugation. Important to note that covariant superderivative involves gauge superfield $ \bar{\fk{A}}_{\tx{super}} $ with bosonic $ \bar{A} $ and fermionic $ \bar{W} $ gauge fields, namely 
	\begin{equation*}\label{key}
		\bar{\sr{D}} = \bar{\partial} +
		i\, \bar{\fk{A}}_{\tx{super}}
		\qquad \quad
		\bar{\fk{A}}_{\tx{super}} = 
		\begin{pmatrix}
			\bar{A} & 0 \\
			\bar{W} & \bar{A}
		\end{pmatrix}
	\end{equation*}
	The model \eqref{Super-CPn} proves to be supersymmetric and integrable, moreover it allows for a further generalisation. From the geometric perspective of the $ \bb{CP}^{n-1} $ model, one of the most nontrivial deformations would involve flowing away from the spherical target space such that the theory remains supersymmetric, integrable and renormalisable. 
	
	\paragraph{Superdeformed $ \bb{CP}^{1} $.} It can be demonstrated that such integrable deformation in the framework above can obtained by deforming one of the currents 
	\begin{equation}\label{Superdeformed_CPn}
		\sr{L}_{\fk{s}} = \sr{V}\bar{\sr{D}}\sr{U} + \bar{\sr{U}}\sr{D}\bar{\sr{V}} + \frac{\varkappa}{2} \, \tx{Tr} \left[ r_{\fk{s}}(\cl{J}) \, \bar{\cl{J}} \right]
	\end{equation} 
	and the $ r_{\fk{s}} $-matrix action can be defined as 
	\begin{equation*}\label{key}
		\begin{array}{c}
			r_{\fk{s}} \left[ \cl{O} \right] = \dfrac{1}{1-\fk{s}}
			\begin{pmatrix}
				\frac{1}{2} (\fk{s} + 1) \cdot \cl{O}_{11} & \sqrt{\fk{s}} \cdot \cl{O}_{12} \\
				\sqrt{\fk{s}} \cdot \cl{O}_{21} & \frac{1}{2} (\fk{s} + 1) \cdot \cl{O}_{22}
			\end{pmatrix}
		\end{array} \,,
	\end{equation*}
	$\fk{s}$ is a deformation parameter. It turns out to be the classical $ r $-matrix satisfying classical Yang-Baxter equation. Such a theory can be shown to be integrable, \textit{i.e.} admitting a Lax pair through $ A_{n} = r_{ns}(\cl{J}) \, d z - r_{ns^{-1}}(\bar{\cl{J}}) \, d \bar{z} $ and zero-curvature condition $ d A_{n} - A_{n} \wedge A_{n} = 0 $, more detailed in \cite{Costello:2017dso,Costello:2018gyb,Costello:2019tri}. However this kind of deformations can break supersymmetry. 
	
	It is clear that the free part of \eqref{Superdeformed_CPn} is super-invariant under \eqref{Worldsheet_Transformations} and it is required to prove that the deformed interaction does so. After resolving all field variations and current differentials it becomes possible to derive the eom of (anti-)holomorphic currents
	\begin{equation}
		\quad \bar{\partial} \cl{J} = \frac{\varkappa}{2} [\cl{J}, r_{s}[\bar{\cl{J}}]] \qquad \partial\bar{\cl{J}} = -\frac{\varkappa}{2} \left[ r_{s}[\cl{J}], \bar{\cl{J}} \right] \,,
	\end{equation}
	so that it can be shown varying interaction leads to 
	\begin{equation}
		\delta \left( \frac{\varkappa}{2} \tx{Tr}(\cl{J} \, r_{s}[\bar{\cl{J}}]) \right) \approx \frac{\varkappa^{2}}{4} \epsilon_{2} \tx{Tr}( r_{s}[\tilde{\cl{J}}] \, [ r_{s}[\cl{J}], \bar{\cl{J}} ])  = \epsilon_{2} \frac{\varkappa^{2}}{4} \tx{Tr}( [r_{s}[\tilde{\cl{J}}] , r_{s}[\cl{J}]] \, \bar{\cl{J}} )
	\end{equation}
	where $\delta \cl{J} = \epsilon_{2} \partial \tilde{\cl{J}} $ and $ \tilde{\cl{J}} = U \otimes B $. By recalling supersymmetry constraints it results in 
	\begin{equation}
		[r_{s}[\tilde{\cl{J}}] , r_{s}[\cl{J}]] = [r_{s}[U \otimes B] , r_{s}[U \otimes V - C \otimes B]] = 0 \,, 
	\end{equation}
	which implies supersymmetry of the novel construction \eqref{Superdeformed_CPn}.

	\section{$ \cl{N} = (2,2) $ K\"ahler sigma model and geometric interpretation} 
	
	It was mentioned in the previous section that this class of models possesses a specific geometric interpretation, which provides deeper understanding of properties, observables and dual maps to other model classes. In order to address this question one can pass to the so called inhomogeneous basis by using field gauge fixing and supersymmetry constraints. For instance, $ U $ and $ C $ can be fixed as $ U_{1} = 1, \, C_{1} = 0 $, which together with $ V \cdot U + B \cdot C = 0 $, $ B \cdot U = 0 $ would provide 
	\begin{equation}
		\begin{aligned}
			U = 
			\begin{pmatrix}
				1 \\
				u
			\end{pmatrix} 
			\quad
			V = 
			\begin{pmatrix}
				-b c -u v \\
				v
			\end{pmatrix}
			\quad
			B = b
			\begin{pmatrix}
				-u \\
				1
			\end{pmatrix}
			\quad
			C = 
			\begin{pmatrix}
				0 \\
				c
			\end{pmatrix} 
		\end{aligned} \,.
	\end{equation}
	Hence the Lagrangian \eqref{Superdeformed_CPn} in this field basis takes the form 
	\begin{equation}
		\sr{L}_{\fk{s}} = v \bar{\partial}u + \bar{u}\partial \bar{v} + b\bar{\partial}c-\bar{c}\partial \bar{b} + \frac{\varkappa \,\fk{s}^{1\over 2}}{1-\fk{s}}\,\left[\alpha |v|^2 + \beta \left(vu\bar{b}\bar{c}+\bar{v}\bar{u} b c\right) + \gamma b c \bar{b}\bar{c}\right]
	\end{equation} 
	where $ \alpha, \, \beta, \, \gamma $ are polynomials in $ u $ and $ \fk{s} $. By extremising the the action w.r.t. $ v $ and resubstituting, one acquires 
	\begin{equation}\label{Kahler_N22_SM}
		\begin{aligned}
			\sr{L}_{\fk{s}} &= \frac{\fk{s}^{-{1\over 2}}-\fk{s}^{1\over 2}}{\alpha \varkappa}\,|\bar{\partial}u|^2 + b \bar{D}c-\bar{c}D \bar{b}+\frac{\varkappa}{\fk{s}^{-{1\over 2}}-\fk{s}^{1\over 2}}\,\left(\gamma-{\beta^2 \over \alpha} |u|^2\right) b  c \bar{b}\bar{c} \, , \\
			& \quad \tx{where} \qquad \bar{D}c=\bar{\partial}c \hspace{-2.5ex} \underbracket[0.1pt][1pt]{ - \,\, {\beta \over \alpha} \bar{u}}_{\Gamma_{uu}^{u} \, \equiv \, \partial_{u} \log g_{u\bar{u}}} \hspace{-2.5ex} \bar{\partial}u \, c \, , \qquad D\bar{b} = \partial\bar{b} \, + \, \underbracket[0.1pt][1pt]{ {\beta \over \alpha} u } \, \partial \bar{u} \, \bar{b} \, .
		\end{aligned}
	\end{equation} 
	In fact, it takes the form of the $ \cl{N}=(2,2) $ K\"ahler $ \sigma $-model \cite{Mirror_Symmetry} 
	\begin{equation}
		\begin{aligned}
			\sr{L}_{\fk{s}} = g_{u\bar{u}} \, |\bar{\partial}u|^2 + b \bar{D}c-\bar{c}D \bar{b} + R_{u\bar{u}u\bar{u}}\,g^{u\bar{u}} g^{u\bar{u}} \, b  c \bar{b}\bar{c} \,,
		\end{aligned}
	\end{equation}
	where $ g $ is the Fateev-Onofri-Zamolodchikov bosonic metric and $ R_{u\bar{u}u\bar{u}}\,g^{u\bar{u}} g^{u\bar{u}} $ is the Ricci scalar in front of the quartic interaction. That completes the mapping between superdeformed chiral formulation \eqref{Superdeformed_CPn} and $ \cl{N}=(2,2) $ $ \sigma $-model \eqref{Kahler_N22_SM}. 
	
	It turns out that the novel construction also has important conformal limits: 
	\begin{itemize}
		\item In the $s \rightarrow 0$ limit it provides a model with $\bb{R} \times S^{1}$ target space and $ \sr{L}_{\mathrm{SCyl}} ={2\over \varkappa}\,\frac{|\bar{\partial}u|^2}{|u|^2}+b \bar{D}c-\bar{c}D \bar{b} $, the so called \textit{super-cylinder}, which after rescaling fermionic dof results in the free theory.
		
		\item In the limit $ u \rightarrow s^{\frac{1}{4}} u $, $ s \rightarrow 0 $ limit, one acquires supersymmetric \textit{cigar} model \cite{WittenCigar,KarchTongTurner}, $ \sr{L}_{\mathrm{SCig}} =\frac{2}{\varkappa}\,\frac{|\bar{\partial}u|^2}{1+|u|^2}+b \bar{D}c-\bar{c}D \bar{b}+{\varkappa \over 2(1+|u|^2)}\,b  c \bar{b}\bar{c} $. 
	\end{itemize} 
	Altogether with limits from the superdeformed chiral side it establishes the following equivalence scheme 
	
	\footnotesize
	\hspace{5.50cm} \textbf{Sigma model} \hspace{1.65cm} \textbf{Chiral model} 
	\vspace{-2mm} 
	\begin{equation}\label{SM_CM_Equivalence}
		\tx{Superdeformed } \bb{CP}^{1} \quad \mapsto \quad 
		\begin{cases}
			u \rightarrow \fk{s}^{\frac{1}{4}} u,\, \fk{s}\rightarrow 0: \, \tx{Supercigar} \,\,\, \leftrightarrow \, \tx{scaled super-GN} \\
			\fk{s}\rightarrow 0: \, \tx{Supercylinder } \mathbb{R} \times S^{1} \, \leftrightarrow \, \tx{Super-Thirring}
		\end{cases}
	\end{equation}
	\normalsize

	\section{2-loop $\beta$-function} 
	
	It can be proven that a new $\bb{CP}$ model is (at least) two-loop renormalisable. Since our superchiral construction can be stated as a 2-dim $\phi^{4}$ field theory analogue, one can investigate the $\beta$-function in terms of perturbation theory. Specifically, one can get it from the corresponding 4-pt function at each order in perturbation, thus taking into account \eqref{Superdeformed_CPn}, for the tree level it is 
	\begin{equation}
		G_4^{\mathrm{tree}}=-\varkappa\,\sum\limits_{a}\,r_{\fk{s}}(\tau_a)\otimes \tau_a \,,
	\end{equation}
	with $\tau_{a}$ to be the underlying algebra generators (to demonstrate, one can restrict to $\fk{sl}_{2}$). For the one-loop one it is required to consider two diagrams, for which it derives 
	\begin{equation}
		G_4^{\mathrm{1\tx{-}loop}}= -{\varkappa^2} \int \,\frac{d^2 z}{(2\pi)^2}\frac{e^{i(p, z)}}{|z|^2}\times {1\over 2}\sum\limits_{a, b}\, [r_{\fk{s}}(\tau_a), r_{\fk{s}}(\tau_b)]  \otimes [\tau_b, \tau_a]
	\end{equation} 
	\begin{equation}
		\mathsf{A}(p):={1\over 2\pi}\int \,d^2 z\frac{e^{i(p, z)}}{|z|^2}=-{1\over 2}\log{\left({p^2 \varepsilon^2}\right)} + \tx{finite} 
	\end{equation} 
	Next we consider the two-loop correction, which accounts for $ 3! $ diagrams 
	\begin{equation}
		\begin{aligned}
			\cl{I}_{4}^{\mathrm{2\tx{-}loop}}& =  -\varkappa^{3}\, e^{i\left(p,z_{13}\right)} \times \frac{r_{\fk{s}}[\tau_a] r_{\fk{s}}[\tau_b] r_{\fk{s}}[\tau_c]}{z_{12} z_{23}} \\ 
			& \otimes \left(\frac{\tau_a \tau_b \tau_c}{\bar{z}_{12} \bar{z}_{23}} + \frac{\tau_a \tau_c \tau_b}{\bar{z}_{13} \bar{z}_{32}} + \frac{\tau_b \tau_a \tau_c}{\bar{z}_{21} \bar{z}_{13}} + \frac{\tau_c \tau_a \tau_b}{\bar{z}_{31} \bar{z}_{12}} + \frac{\tau_b \tau_c \tau_a}{\bar{z}_{23} \bar{z}_{31}} + \frac{\tau_c \tau_b \tau_a}{\bar{z}_{32} \bar{z}_{21}}\right)
		\end{aligned} \,,
	\end{equation} 
	
	\vspace{-5.00mm}
	\begin{figure}[!h]
		\centering
		\includegraphics[width=0.50\textwidth]{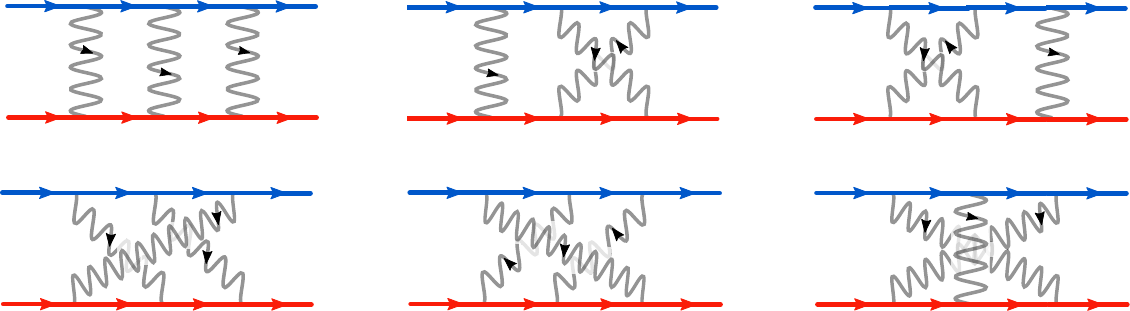}
	\end{figure}
	\nd where $z_{ij} = z_{i} - z_{j}$ comes from the field lines. So that the two-loop contribution can be given as 
	\begin{equation}
		\begin{aligned}
			&G_4^{\mathrm{2\tx{-}loop}} = -\varkappa^3\, \int \,\frac{d^2 z_{12}}{(2\pi)^2}\,\frac{d^2 z_{23}}{(2\pi)^2}\frac{e^{i(p, z_{13})}}{z_{12}z_{23}}\,\sum\limits_{a, b, c}\,r_{\fk{s}}(\tau_a)r_{\fk{s}}(\tau_b)r_{\fk{s}}(\tau_c) \\ 
			& \hspace{2cm}\otimes\left({1\over \bar{z}_{12}\bar{z}_{23}}\,[\tau_a, [\tau_b, \tau_c]]-{1\over \bar{z}_{12}\bar{z}_{13}}\,[\tau_b, [\tau_a, \tau_c]]\right) 
		\end{aligned} \,.
	\end{equation} 
	From here it can be further proven that both terms above are proportional to $A^{2}$ 
	\begin{equation}
		\begin{aligned}
			& G_4^{2\tx{-}loop} = -\dfrac{\varkappa^3}{(2\pi)^2} \, A(p)^2\,\sum\limits_{a, b, c} \, r_{\fk{s}}(\tau_a)r_{\fk{s}}(\tau_b)r_{\fk{s}}(\tau_c) \\
			& \otimes \left(\,[\tau_a, [\tau_b, \tau_c]]-{1\over 2}\,[\tau_b, [\tau_a, \tau_c]]\right) + \tx{finite}
		\end{aligned} 
		\,\rightarrow \,
		\begin{aligned}
			G_4^{\mathrm{2\tx{-}loop}} = -{\varkappa^3\over 2\pi^2} \, A(p)^2\,\,\sum\limits_{a}\,\ddot{r}_{\fk{s}}(\tau_a) \otimes \tau_a \,,
		\end{aligned}
	\end{equation}
	where on the right hand side we assumed the $\fk{sl}_{2}$ case. Important to note that in the one-loop derivation the renormalisability was consistent with the obtained Nahm-type \cite{Nahm} constraint 
	\begin{equation}
		\dot{r}_{\fk{s}}([\mathbb{A}, \mathbb{B}])=[r_{\fk{s}}(\mathbb{A}), r_{\fk{s}}(\mathbb{B})] \,,
	\end{equation}
	where $\bb{A,B}$ are algebra generators and $ \dot{r}_{\fk{s}} = \fk{s} \, d_{\fk{s}} r_{\fk{s}} $. This condition also independently implied supersymmetric invariance of the model.

	\section{Duality: Supercylinder/Super-Thirring} 
	
	\paragraph{Super-Thirring.} To provide explicit confirmation of the mapping equivalence \eqref{SM_CM_Equivalence}, we shall  provide the computation of the 4-pt function on both sides of the Supercylinder/Super-Thirring (SC/ST) \cite{Freedman_CPSSGTM,Freedman_SGITA} correspondence. From the ST it implies perturbation theory in $ \varkappa $ and from SC computation is done through the associated vertex operators. Hence from the ST side for the bosonic\footnote{Including fermions leads to additional Koba-Nielsen factors} 4-pt correlator $\Gamma_{4} = \langle u(\tilde{z}_{1})v(\tilde{z}_{2}) \bar{u}(z_{1})\bar{v}(z_{2}) \rangle$ at tree level one obtains 
	\small
	\begin{equation}
		\begin{aligned}
			& I_0(\tilde{z}_1, \tilde{z}_2 | z_1, z_2) = - {\varkappa\over (2\pi)^2}\,\int\,{d^2 z \over (2\pi)^2} \,\frac{1}{(\tilde{z}_1-z)(z-\tilde{z}_2)(\bar{z}_1-\bar{z})(\bar{z}-\bar{z}_2)} \\ 
			& \quad = -\dfrac{1}{(2 \pi)^{3}}\frac{\varkappa}{2 \, \tilde{z}_{12}\bar{z}_{21}} \,\, \log{[\mathsf{CR}(\tilde{z}_1, \tilde{z}_2| z_1, z_2)]} \,,
		\end{aligned}
	\end{equation}
	\normalsize
	where $ \mathsf{CR} $ denotes $\tx{SL}(2,\bb{C})$ invariant cross-ratio. By virtue of induction for the $ (\ell+1) $-loop it becomes 
	\small
	\begin{equation}
		\begin{aligned} 
			& I_{\ell+1}=-{\varkappa^{\ell+2}\over  (2\pi)^2}\,\sum\limits_{i=0}^{\ell + 1}\,\sum\limits_{p\in S_{\ell+1}}\!\!\int\,{d^2w \over (2\pi)^2}\,\prod\limits_{i=1}^{\ell + 1}\,{d^2 w_i\over (2\pi)^2}\,\frac{1}{(\tilde{z}_1-w)(w-w_1)\cdots (w_{\ell + 1}-\tilde{z}_2)}\times \\ 
			& \hspace{2.5cm}\, \times\frac{1}{(\bar{z}_1-\bar{w}_{1'})\cdots (\bar{w}_{i'}-\bar{w})(\bar{w}-\bar{w}_{i+1'})\cdots (\bar{w}_{\ell + 1'}-\bar{z}_2)}
		\end{aligned}
	\end{equation}
	\normalsize
	\vspace{-4.00mm}
	\begin{figure}[!h]
		\centering
		\includegraphics[width=0.45\textwidth]{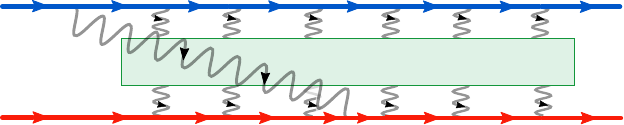} 
		\caption*{\footnotesize Fig.: Due to internal loop cancellations only ladder diagrams appear (internal box implies all possible permutations at a given loop level). \normalsize}
	\end{figure} 
	
	\vspace{-5.00mm}
	\nd It is then possible to deduce a recursion in terms lower loop integrals and resolve it, which leads to
	\begin{equation}
		I_{\ell}(\tilde{z}_1, \tilde{z}_2 | z_1, z_2)={1\over (2\pi)^2}\frac{1}{\bar{z}_{21} \tilde{z}_{12}}\,\frac{1}{(\ell+1)!}\,\left(-{\varkappa\over 4\pi}\right)^{\ell+1}\left[\log\,\mathsf{CR}(\tilde{z}_1, \tilde{z}_2| z_1, z_2)  \right]^{\ell+1} \,.
	\end{equation} 
	Resumming all loop contributions results in a 4-pt correlator 
	\begin{equation}\label{4pt_correlator_ST}
		\Gamma_{4}=\sum\limits_{\ell=-1}^\infty I_{\ell}={1\over (2\pi)^2}\frac{1}{\bar{z}_{21} \tilde{z}_{12}}\,\left[\mathsf{CR}(\tilde{z}_1, \tilde{z}_2| z_1, z_2) \right]^{-{\varkappa\over 4\pi}} \,.
	\end{equation}
	
	\paragraph{Supercylinder.} From the worldsheet perspective one can rewrite the SC Lagrangian in the form 
	\begin{equation}
		\mathcal{L} = {2\over \varkappa} \frac{|\bar{\partial}u|^2}{|u|^2}+B\bar{\partial}C-\bar{C} \partial \bar{B} \rightarrow \mathcal{L}={2\over \varkappa}\, |\bar{\partial}W|^2 + i\,\bar{\Psi} \slashed{\partial}\Psi \quad \Psi = \begin{pmatrix} B \\ \bar{C} \end{pmatrix} \,,
	\end{equation} 
	which in terms of vertex operators maps to elementary fields 
	\small 
	\begin{equation}
		u=e^{W}\,,\quad\quad v=\left({2\over \varkappa}\,\partial\bar{W}-BC\right) \,e^{-W}\,,\quad\quad b=B\,e^{-W}\,,\quad\quad c=C\,e^{W}
	\end{equation}
	\normalsize 
	The corresponding 4-pt function can be given by the following path integral 
	\begin{equation}
		\begin{aligned}
			& \langle u(\tilde{z}_1)\,v(\tilde{z}_2)\,\bar{u}(z_1)\,\bar{v}(z_2)\rangle = \\ 
			& = \int\,\left({2\over \varkappa}\,\partial \bar{W}(\tilde{z}_2)-B(\tilde{z}_2)C(\tilde{z}_2)\right) \left(-{2\over \varkappa}\,\bar{\partial} W(z_2)-\bar{B}(z_2)\bar{C}(z_2)\right)\, \\ 
			& \times \exp{\left[-\,\int \!i\,dz\!\!\wedge \!\!d\bar{z}\;\mathcal{L}+W(\tilde{z}_1) - W(\tilde{z}_2) + \bar{W}(z_1) - \bar{W}(z_2)\right]}
		\end{aligned}
	\end{equation}
	After substituting $ W = {\varkappa\over 4\pi}\,\log{\left({|z-z_2|^2\over|z-z_1|^2}\right)} $ and $ \bar{W}={\varkappa\over 4\pi}\,\log{\left({|z-\tilde{z}_2|^2 \over|z-\tilde{z}_1|^2}\right)} $ eom, it derives for the 4-pts 
	\begin{equation}\label{4pt_correlator_SC}
		\langle u(\tilde{z}_1)\,v(\tilde{z}_2)\,\bar{u}(z_1)\,\bar{v}(z_2)\rangle={1\over (2\pi)^2}\frac{1}{\bar{z}_{21} \tilde{z}_{12}}\,\,\left[\mathsf{CR}(\tilde{z}_1, \tilde{z}_2| z_1, z_2) \right]^{-{\varkappa\over 4\pi}} \,.
	\end{equation} 
	It proves complete agreement of Supercylinder \eqref{4pt_correlator_SC} and Super-Thirring \eqref{4pt_correlator_ST}. 
	
	\vspace{-5.00mm}
	\paragraph{Remarks.} The new superdeformed class of models establishes important relations between chiral and sigma models and their conformal limits. It would be important to investigate its further mapping with $A$- and $B$-model (Landau-Ginzburg class) sectors and $\cl{N=2}$ sine-Liouville theory in particular. The last would allow to solve the novel $\bb{CP}$ class by virtue of Thermodynamic Bethe Ansatz \cite{Fendley:1992dm,Fendley:1993pi} or the generalised fermionisation procedure \cite{Wiegmann:1985jt}. It is of current investigation the emergence of new class from the 4-dim Chern-Simons theory with $\bb{CP}$ order defects and its IR discretisation \cite{Four-dim_CS_OD}. Further questions can include finding complete GLSM formulation and computations of instanton corrections. 
	
	\vspace{-5.00mm}
	\paragraph{Acknowledgment.} I would like to thank C. Ahn, Z. Bajnok, D. Gaiotto, B. Vicedo, C. Klimčík, A. Losev, A. Tseytlin and A. Smilga for discussions, and especially P. Fendley and M. Roček for comments on the manuscript. A. P. also expresses gratitude for the support from Russian Science Foundation grant \href{https://rscf.ru/en/project/22-72-10122/}{RSCF-22-72-10122} and European Union – NextGenerationEU, from the program STARS@UNIPD, under project "Exact-Holography -- A new exact approach to holography: harnessing the power of string theory, conformal field theory, and integrable models".

\newpage
\bibliographystyle{JHEP}
\bibliography{SDSM}


\end{document}